# Outage Probability Analysis of Selective-Decode and Forward Cooperative Wireless Network over Time Varying Fading Channels with Node Mobility and Imperfect CSI Condition

Ravi Shankar
*Department of ECE*
*NIT Patna*
Patna, India
ravi.mrce@gmail.com

Ritesh Kumar Mishra
*Department of ECE*
*NIT Patna*
Patna, India
ritesh@nitp.ac.in

*Abstract*— In this work, we explore the outage probability (OP) analysis of selective decode and forward (S-DF) cooperation protocol employing multiple-input multiple-output (MIMO) orthogonal space-time block-code (OSTBC) over time varying Rayleigh fading channel conditions with imperfect channel state information (CSI) and mobile nodes. The closed-form expressions of the per-block average OP, probability distribution function (PDF) of sum of independent and identically distributed (i.i.d.) Gamma random variables (RVs), and cumulative distribution function (CDF) are derived and used to investigate the performance of the relaying network. A mathematical framework is developed to derive the optimal source-relay power allocation factors. It is shown that source node mobility affects the per-block average OP performance more significantly than the destination node mobility. Nevertheless, in other node mobility situations, cooperative systems are constrained by an error floor with a higher signal to noise ratio (SNR) regimes. Simulation results show that the equal power allocation is the only possible optimal solution when source to relay link is stronger than the relay to destination link. Also, we allocate almost all the power to the source node when source to relay link is weaker than the relay to destination link. Simulation results also show that OP simulated plots are in close agreement with the OP analytic plots at high SNR regimes.

## I. INTRODUCTION

Spatial diversity is the main advantages of MIMO systems [1]-[2]. Despite of advantages, antenna diversity may not be practical for many wireless scenarios due to hardware limitations, cost and size of the equipment. Also, non-cooperative transmission is very expensive due to high transmit power necessary for reliability of the wireless system. This might happen, for example, in a multiple hop wireless system where users have power constraint for signal transmission. In addition, relay dependent cooperative communication can likewise improve the footprint of traditional wireless network by utilizing the previously unexploited user-equipment's (UEs) of other cellular nodes as relaying nodes [3]. Cooperative diversity found its way in Next generation commercial wireless technologies, e.g., LTE advances and fifth generation wireless technology (5G) [4] because by using relaying, we can get a data transmission rate of 150 Mbps in the downlink (D/L) and up to 350 Mbps in the uplink (U/L) with lower battery power consumption.

Relaying protocol improves the network's spectral efficiency (bits/second/Hertz) by forming a virtual array of multiple antennas. Relaying network generally can be divided into two categories, namely, amplify-and-forward (AF) [5] and decode-and-forward (DF) [6], including S-DF [7], incremental DF [8], and so on. The DF relaying protocol decodes and re-encodes the received signal from source node, then it forwards it to the destination node. The disadvantage of this protocol is that it may send incorrectly decoded signal to the receiving terminal.

In work [9], the authors investigated the symbol error rate (SER) performance of multiple relay AF relaying network in the presence of node mobility and imperfect CSI conditions. Two cooperation techniques are considered; best relay selection (BRS) and conventional cooperation protocol. Results show that BRS protocol outperforms the conventional relay protocol. However, the main disadvantage of the above works is noise amplification due to AF protocol.

To overcome the drawback of DF and AF relaying protocol S-DF protocol is introduced in which relay forwards only correctly decoded signal to the destination. In the works [10]-[13], the authors investigated the S-DF relaying protocol over time invariant Nakagami-m fading channel conditions. But these works have not considered the real time communication constraints such as time varying channel links due to node mobility, Doppler spread, keyhole effect and imperfect CSI.

In the work [14]-[16], the authors investigated the PEP performance of multiple relay S-DF protocol over time varying Raleigh fading channel conditions. These works investigated the effect of time-varying fading arising due to mobility of nodes and imperfect CSI on the PEP performance of multiple relay multiple phase multiple hop MIMO STBC multiple relay relaying network. A mathematical framework is derived for evaluating the optimal power allocation factors which enhance the PEP performance of the system. Up to now, to the best of our knowledge, there has been no work that has analyzed the outage performance of MIMO STBC S-DF over time varying fading channel conditions.

The remainder of this paper is organized as follows: In section II we have given system model for multiple relay

S-DF cooperative communication system. Section III presents an investigation of the OP and derivation of optimal power factors for multiple relay dual hop dual phase S-DF cooperation protocol. Simulation result is given in section IV. Lastly, paper is concluded in section V.

## II. MULTIPLE RELAY DUAL HOP DUAL PHASE S-DF PROTOCOL SYSTEM MODEL

Consider MIMO S-DF relaying network in which source, relay and destination nodes are equipped with $N_S, N_R$ and $N_D$ number of antennas. The Schematic representation of dual hop dual phase multiple relay S-DF cooperation system is given in Fig. 1. Since we employ same OSTBC code at source and relay node so $N_R = N_S = N$ [14]-[15]. The signal transmission is divided into two phases. In the first phase of signal transmission OSTBC code $B_S(\Upsilon) \in \mathbb{C}^{N \times T_S}$ is transmitted from the source to destination and all relay nodes, where $T_S$ and $\Upsilon$ defines the number of time slots and $\Upsilon^{th}$ code-word in a block of $M_b$ code-word matrices respectively. The received code-word's at the destination and $r^{th}$ relay are given as [14]-[15],

$$R_{SD}(\Upsilon) = \sqrt{\frac{P_S}{NR_C}} \mathbb{R}_{SD}(\Upsilon) B_S(\Upsilon) + Z_{SD}(\Upsilon) \quad (1)$$

$$R_{SR}^{(r)}(\Upsilon) = \sqrt{\frac{P_S}{NR_C}} \mathbb{R}_{SR}^{(r)}(\Upsilon) B_S(\Upsilon) + Z_{SR}^{(r)}(\Upsilon); \quad 1 \leq r \leq L. \quad (2)$$

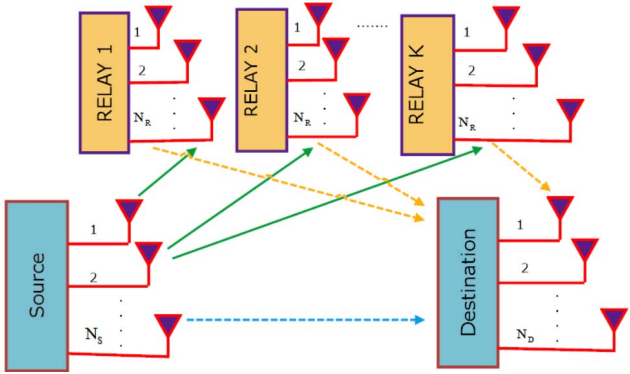

Fig. 1. Schematic representation of dual hop dual phase multiple relay S-DF cooperation system.

Where $P_S$ and $R_C$ denotes the total available power at the source node and code-rate of OSTBC code respectively. The terns of source-destination $\mathbb{R}_{SD}(\Upsilon) \in \mathbb{C}^{N_D \times N}$ and source - $r^{th}$ relay $\mathbb{R}_{SR}^{(r)}(\Upsilon) \in \mathbb{C}^{N \times N}$ matrices are presented as $d^{SD}[n_1, m_1]$ and $d^{SR}[l_1, m_1]$ respectively. These terms are Gaussian distributed, circular shift with zero average value and average gains $\Omega_{SD}^2$ and $(\Omega_{SR}^{(r)})^2$ respectively. In the second phase of signal transmission, $r^{th}$ relay node forwards only correctly decoded signal to the destination node, otherwise it will remain in the idle state. The received codeword symbol at the destination node due to transmission from $r^{th}$ relay nodes is given as [14]-[15],

$$R_{RD}^{(r)}(\Upsilon) = \sqrt{\frac{P_r}{NR_c}} \mathbb{R}_{RD}^{(r)}(\Upsilon) B_S(\Upsilon) + Z_{RD}^{(r)}(\Upsilon) \quad (3)$$

The terns of $r^{th}$ relay-destination $\mathbb{R}_{RD}^{(r)}(\Upsilon) \in \mathbb{C}^{N_D \times N}$ matrix are presented $d^{RD}[l_1, n_1]$. These terms are Gaussian distributed, circular shift with zero average value and average gains $(\Omega_{RD}^{(r)})^2$ respectively. The noise matrices $Z_{SD} \in \mathbb{C}^{N_D \times T_S}$, $Z_{SR}^{(r)}(\Upsilon) \in \mathbb{C}^{N \times T_S}$ and $Z_{RD}^{(r)}(\Upsilon) \in \mathbb{C}^{N_D \times T_S}$ are Gaussian distributed with zero average value and average gain is $N_0 / 2$ (per complex dimension). Due to node mobility fading channels are time selective in nature, modeled as first order autoregressive model, given below [14]-[15],

$$\mathbb{R}_i(\Upsilon) = \varepsilon_i \mathbb{R}_i(\Upsilon - 1) + \sqrt{1 - \varepsilon_i^2} \mathbb{Q}_i(\Upsilon); \\ i \in \{SD, SR, RD\}. \quad (4)$$

The source-destination, source- $r^{th}$ relay and $r^{th}$ relay-destination fading link's correlation coefficients are denoted as $\varepsilon_{SD}$, $\varepsilon_{SR}$ and $\varepsilon_{RD}$ respectively. We can evaluate the these correlation coefficients by using Jake's classical wireless propagation channel model [1], [2], [15], given as, $\varepsilon = J_0(2\pi f_C v_p / R_S C_P)$. Where $f_C, v_p, R_S$ and $C_p$ denote the carrier frequency, relative velocity, symbol transmission rate and velocity of light respectively. The source to destination, source to $r^{th}$ relay and $r^{th}$ relay to destination fading link's time varying matrices are denoted as $\mathbb{Q}_{SD}, \mathbb{Q}_{SR}$ and $\mathbb{Q}_{RD}$ respectively. These matrices are distributed as complex Gaussian noise with zero average value and variances $\sigma_e^{SD}, \sigma_e^{SR}$ and $\sigma_e^{RD}$ respectively.

## III. OUTAGE PROBABILITY ANALYSIS OF DUAL PHASE DUAL HOP MULTIPLE RELAY S-DF PROTOCOL

### A. Outage Probability derivation

For multiple relay dual hop S-DF protocol, we can write the expression of outage probability $\overline{P}_{outage}(\gamma_0)$ for multiple relay scenario as [15],

$$\overline{P}_{outage}(\gamma_0) = \frac{1}{M_b} \sum_{\Upsilon=1}^{M_b} \left[ \left( P^{SR}(\gamma_0) \right) \times P^{SD-RD}(\gamma_0) \right]. \quad (5)$$

Where $P^{SR}(\gamma_0)$ and $P^{SD-RD}(\gamma_0)$ denote the outage probability for source to $r^{th}$ relay and source- $r^{th}$ relay-destination channel link respectively, expressed in (6) and (7) respectively.

$$P^{SR}(\gamma_0) = \prod_{r \in \overline{\psi}} \left\{ P_R(\gamma_{SR_r}(\Upsilon) \leq \gamma_0) \right\} \quad (6)$$

$$P^{SD-RD}(\gamma_0) = P_R \left\{ (\gamma_{SD}(\Upsilon) + \sum_{r \in \psi} \gamma_{R_r D}(\Upsilon)) \leq \gamma_0 \right\}. \quad (7)$$

Where sets $\psi$ and $\overline{\psi}$ includes the correctly and incorrectly decoded relays respectively, and $\gamma_0$ denotes the outage threshold. Following [14], [15], [16] we can write an expression $P^{SR}(\gamma_0)$ in terms of incomplete gamma function as [15],

$$P^{SR}(\gamma_0) = \prod_{r \in \bar{\psi}} \left\{ \frac{\gamma(N^2, \frac{\gamma_0}{M_{SR}^{(r)}(\tilde{\Omega}_{SR}^{(r)})^2})}{(N^2-1)!} \right\}. \quad (8)$$

Where $\gamma(.,.)$ denotes the lower incomplete Gamma function [1], [14]-[15] $(\tilde{\Omega}_{SR}^{(r)})^2 = (\Omega_{SR}^{(r)})^2 + (\sigma_\in^{SR})^2$ and $M_{SR}^{(r)}$ is expressed in (9) [15]. Where $N_a$ denotes the number of non-zero symbol transmissions per codeword, $(\sigma_\in^{SR})^2$ denotes the channel error variance for the source to $r^{th}$ relay link. Also to find out the formulation of $P_{SD-RD}(\gamma_0)$ we have to use the concept of the sum of i.i.d. Gamma RV's [14], [15], [16]. Following (18), $P^{SD-RD}(\gamma_0)$ can be expressed in (10). Where in expression (10) we consider that the set $\psi$ contain values $\psi = [1,2,3,4,...,k]$.

Where $\alpha_i = NN_D$, $\beta_1 = 1/\{M_{SD}((\Omega_{SD})^2 + (\sigma_\in^{SD})^2)\}$, $\beta_2 = \beta_2 = ....... = \beta_L = 1/\{M_{RD}^{(r)}((\Omega_{RD}^{(r)})^2 + (\sigma_\in^{RD})^2)\}$

$$M_{SD} = \frac{(P_S/N_0 N R_C)\varepsilon_{SD}^{2(\Upsilon-1)}}{(1+(P_S/N_0 N R_C)\varepsilon_{SD}^{2(\Upsilon-1)} N_a(\sigma_\in^{SD})^2 + (P_S/N_0 N R_C)(1-\varepsilon_{SD}^{2(\Upsilon-1)})N_a(\sigma_e^{SD})^2)}.$$

Where $(\sigma_\in^{RD})^2$ and $(\sigma_\in^{SD})^2$ denote the channel error variance of the $r^{th}$ relay to the destination and source to destination links respectively. Substituting (8) and (10) into (7) we can express outage probability in (11).

### B. Optimal power allocation analysis of dual phase dual hop multiple relay S-DF protocol

For optimal source to relay power allocation we consider perfect CSI and static conditions, i.e.,
$(\sigma_\in^{SR})^2 = (\sigma_\in^{SD})^2 = (\sigma_\in^{RD})^2 = 0$ and $(\sigma_e^{SR})^2 = (\sigma_e^{SD})^2 = (\sigma_e^{RD})^2 = 0$.

Using the identity $\gamma(s_1, x_1) = x_1^{s_1} \sum_{l_1=0}^{\infty} \frac{(-x_1)^{l_1}}{l_1!(s_1+l_1)}$

and $_1F_1(\alpha';\beta';z') = \sum_{n'=0}^{\infty} \frac{(\alpha')^{(n')}(z')^{n'}}{(\beta')^{(n')}n'!}$ and considering the terms corresponding to $l_1 = 0$ and $n' = 1$ the outage probability expression for $L = 2$ given in (11) can be expressed in (12). We can formulate the outage probability expression in terms of convex optimization [15], [16] problem for finding the values of optimal power allocation factors $\beta_0$, $\beta_1$ and $\beta_2$ as given in (13). Where we consider $(\Omega_{SR}^{(1)})^2 = (\Omega_{SR}^{(2)})^2 = (\Omega_{SR})^2$ and $(\Omega_{RD}^{(1)})^2 = (\Omega_{RD}^{(2)})^2 = (\Omega_{RD})^2$ for deriving the optimal source to relay power allocation factors.

### IV. SIMULATION RESULT

In this section, we present the impact of mobile environment and imperfect channel estimation on the per block average OP performance for S-DF relaying network and furthermore support the theoretical results presented in section III. Fig. 2 shows that due to mobility and time selective fading channel conditions, per block average PEP performance of the wireless system degrades in comparison to the static and perfect CSI conditions, i.e., $\varepsilon_i = 1; i \in \{SD, SR, RD\}$. Additionally, the low estimation of correlation coefficients at high velocity yields severe degradation in the PEP performance [15]. Further, the wireless system provides a decent antenna diversity $NN_D + NL\min(N,N_D)$ for static nodes and perfect CSI.

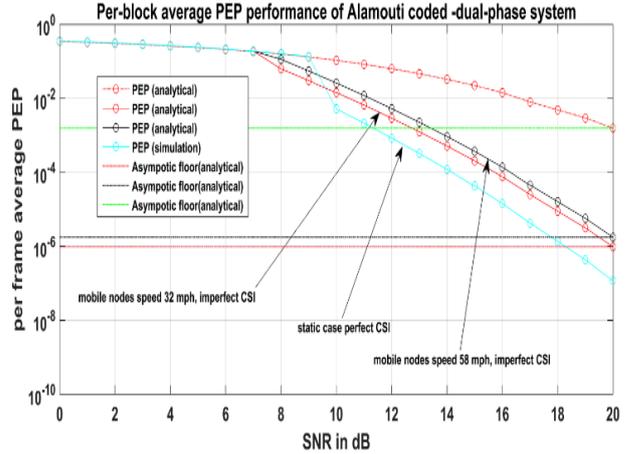

Fig. 2. Per-block average PEP versus SNR in dB for multiple relay dual hop dual phase S-DF protocol with 4-PSK Alamouti-coded code-word, $f_c = 5.9$ GHz (ITS band), $R_S = 10\,kbps$, $N_0 = 1, M_B = 20$, and $\varepsilon_i = \{0.9915, 0.9724\}, v_p = \{32,58\}\,mi/h, \sigma_{\in_i}^2 = 0.01$ $\sigma_{e_i}^2 = 0.10, N = N_D = 2, R_C = 1, L = 2, \Omega_{SD}^2 = \Omega_{SR}^2 = \Omega_{RD}^2 = 2.$

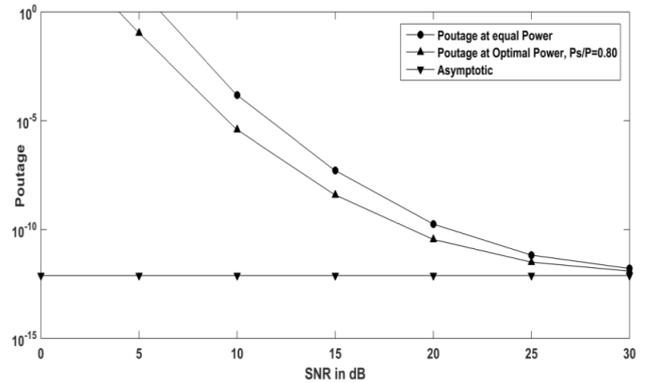

Fig. 3. Per-block average OP versus SNR in dB for multiple relay dual hop dual phase S-DF protocol with 4-PSK Alamouti-coded code-word, $f_c = 5.9\,GHz(ITS\,Band), R_S = 10\,kbps, N_0 = 1, M_B = 15$, and $\varepsilon_i = 0.9915, v_p = 32\,mi/h, \sigma_{\in_i}^2 = 0.01, v_p = 32\,mi/h, \sigma_{e_i}^2 = 0.10,$ $\sigma_{e_i}^2 = 0.10, N = N_D = 2, R_C = 1, L = 2, \Omega_{SD}^2 = \Omega_{SR}^2 = \Omega_{RD}^2 = 2.$

It can be seen from fig. 3 that the OP simulated plots are in close agreement with the OP analytic plots at high SNR regimes. One can also show that the optimal power allocation factors evaluated utilizing (13) enhance the OP performance significantly in the low and medium SNR regimes. Further, at high SNR, the OP performance for both equal and optimal power allocation approaches the derived asymptotic limit. Fig. 4 demonstrates the theoretical and simulated PEP probability graphs considering all the scenarios, i.e., source to destination, source to relay and relay to destination links have the same channel variances.

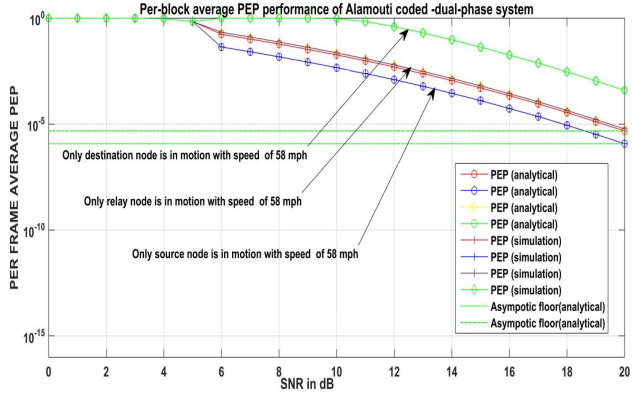

Fig. 4. Per-block average PEP versus SNR in dB for multiple relay dual hop dual phase S-DF protocol with 4-PSK Alamouti-coded code-word, $f_c$ = 5.9 GHz (ITS band), $R_S = 11\ kbps$, $N_0 = 1$, $M_B = 10$ and $\varepsilon_i = \{0.9724\}$, $v_p = 58\ mi/h$, $\sigma^2_{\in_i} = 0.01$, $\sigma^2_{e_i} = 0.10$, $N = N_D = 2$, $R_C = 1$, $L = 2$, $\Omega^2_{SD} = \Omega^2_{SR} = \Omega^2_{RD} = 2$.

## V. CONCLUSION

In this work, we investigate the OP performance of dual phase dual hop multiple relay cooperation protocol. A mathematical framework is developed to derive the optimal source-relay power allocation factors which significantly enhance the system performance. Also, we find that node mobility yields time selective fading, which degrades the OP performance in contrasted with the static environment.

## ACKNOWLEDGMENT

The author would like to acknowledge the Ministry of Electronics & Information Technology (MeitY), Government of India for supporting the financial assistant during research work through "Visvesvaraya PhD Scheme for Electronics & IT".

## APPENDIX

Let $\{\phi_l\}_{l=1}^{N}$ be independent Gamma random variables with shape and rate parameters $\alpha_l$ and $\beta_l$ respectively. Then the PDF of $K = \phi_1 + \phi_2 + ... + \phi_N$ is expressed in (14). Where $\beta_1 = \min_l\{\beta_l\}$, $A = \prod_{i=1}^{N}\left(\frac{\beta_1}{\beta_i}\right)^{\alpha_i}$ and the coefficient $\delta_l$ can be obtained from the recursive relations as given below

$$\begin{cases} \delta_0 = 1 \\ \delta_{l+1} = \frac{1}{l+1}\sum_{i=1}^{l+1}\left(\sum_{j=1}^{N}\alpha_j\left(1-\frac{\beta_1}{\beta_j}\right)^i\right)\delta_{l+1-i}; \quad l = 0, 1, 2,... \end{cases}$$

(15)

The Moschopoulos PDF representation [1], [2] of the summation of Gamma variates can be applied to any $\{\alpha_l\}_{l=1}^{N}$ i.e., some $\{\alpha_l\}_{l=1}^{N}$ can be distinct while others can be equal. The cumulative distribution function (CDF) of $K$ can be derived from the PDF as,

$$F_K(\xi) = A\sum_{n=0}^{\infty}\delta_n\int_0^{\xi}\frac{\xi^{\sum_{i=1}^{N}\alpha_i+n-1}e^{-\xi/\beta_1}}{\left(\sum_{i=1}^{N}\alpha_i+n-1\right)!\beta_1^{\sum_{i=1}^{N}\alpha_i+n}}d\xi$$

(16)

CDF expression given above can be simplified using the outcomes of [1],[2], i.e.,

$$\int_0^u \xi^{v-1}e^{-\mu\xi}d\xi = \mu^{-v}\Psi(v, \mu\xi) \quad \text{for } Real(v) > 0.$$

Where $\Psi(.,.)$ denotes the lower incomplete Gamma [15] function expressed as: $\Psi(\kappa, \xi) = \int_0^{\xi}e^{-t}t^{\kappa-1}dt$.

Therefore,

$$F_K(\xi) = A\sum_{n=0}^{\infty}\delta_n\frac{\beta_1^{(\sum_{i=1}^{N}\alpha_i+n)}}{\left(\sum_{i=1}^{N}\alpha_i+n-1\right)!\beta_1^{\sum_{i=1}^{N}\alpha_i+n}}\times\Psi\left(\sum_{i=1}^{N}\alpha_i+n, \xi/\beta_1\right)$$

(17)

We can also express (16) in terms of confluent hypergeometric series as given in (18). Where $_1F_1(.;.;.)$ is the confluent hypergeometric function [1], [2], [14], [15]. We have used the expression $\Psi(\Im,\xi) = \Im^{-1}\xi^{\Im}{}_1F_1(\Im; 1+\Im; -\xi)$ for deriving the result given in (17). We can find out the value of lower incomplete Gamma function and the confluent hypergeometric function by using the software such as MATLAB and MATHEMATICA.

$$\mathrm{M}_{SR}^{(r)} = \frac{(P_S/N_0 NR_C)\varepsilon_{SR}^{2(\Upsilon-1)}}{(1+(P_S/N_0 NR_C)\varepsilon_{SR}^{2(\Upsilon-1)} N_a (\sigma_\epsilon^{SR})^2 + (P_S/N_0 NR_C)(1-\varepsilon_{SR}^{2(\Upsilon-1)}) N_a (\sigma_e^{SR})^2)} \tag{9}$$

$$P^{SD-RD}(\gamma_0) = \prod_{i=1}^{k}\left(\frac{\beta_1}{\beta_i}\right)^{\alpha_i} \sum_{n=0}^{\infty} \delta_n \frac{\left(\sum_{i=0}^{k}\alpha_i+n\right)^{-1} \gamma_0^{\left(\sum_{i=0}^{k}\alpha_i+n\right)}}{\left(\sum_{i=0}^{k}\alpha_i+n-1\right)! \beta_1^{\sum_{i=0}^{k}\alpha_i+n}} \times {}_1F_1\left(\sum_{i=0}^{k}\alpha_i+n;\ 1+\sum_{i=0}^{k}\alpha_i+n;\ \frac{-\gamma_0}{\beta_1}\right). \tag{10}$$

$$\overline{P}_{outage}(\gamma_0) = \frac{1}{M_b}\sum_{\Upsilon=1}^{M_b}\left[\prod_{r\in\overline{\psi}}\left\{\gamma(N^2,\frac{\gamma_0}{\mathrm{M}_{SR}^{(r)}(\tilde{\Omega}_{SR}^{(r)})^2})\right\}\times \prod_{i=1}^{k}\left(\frac{\beta_1}{\beta_i}\right)^{\alpha_i}\sum_{n=0}^{\infty}\delta_n \frac{\left(\sum_{i=0}^{k}\alpha_i+n\right)^{-1}\gamma_0^{\left(\sum_{i=0}^{k}\alpha_i+n\right)}}{\left(\sum_{i=0}^{k}\alpha_i+n-1\right)!\beta_1^{\sum_{i=0}^{k}\alpha_i+n}} \\ \times {}_1F_1\left(\sum_{i=0}^{k}\alpha_i+n;\ 1+\sum_{i=0}^{k}\alpha_i+n;\ \frac{-\gamma_0}{\beta_1}\right)\right]. \tag{11}$$

$$\overline{P}_{outage}(\gamma_0) = \frac{1}{M_b}\sum_{k=1}^{M_b}\left[\begin{array}{l}(\gamma_0 NR_C)^{2N^2+NN_D}\left(\frac{1}{\Omega_{SR}^2}\right)^{2N^2}\left(\frac{1}{\Omega_{SD}^2}\right)^{NN_D}(1/N^2!)^2 \\ \times(1/NN_D)(N_0/P)^{2N^2+NN_D}(1/\beta_0)^{2N^2+NN_D} \\ +2(\gamma_0)^{N^2+NN_D}(NR_C)^{N^2+4NN_D}\left(\frac{1}{\Omega_{SR}^2}\right)^{N^2}\left(\frac{1}{\Omega_{SD}^2}\right)^{NN_D}\left(\frac{1}{\Omega_{RD}^2}\right)^{2NN_D} \\ \times(N_0/P)^{N^2+4NN_D}(1/\beta_0)^{2N^2+NN_D}(1/N^2!)(1/(2NN_D-1)!)(1/\beta_0)^{N^2+2NN_D} \\ \times(1/\beta_2)^{2NN_D} + \frac{(\gamma_0)^{3NN_D-1}}{(4P^2)^{3NN_D}(\beta_0\beta_1)^{3NN_D}(3NN_D-1)!}\end{array}\right] \tag{12}$$

$$\min_{\beta_0, \beta_1, \beta_2} \left[ \begin{array}{l} K_1(1/\beta_0)^{2N^2+NN_D} + K_2(1/\beta_0)^{2N^2+NN_D}(1/\beta_0)^{N^2+2NN_D} \times (1/\beta_2)^{2NN_D} \\ +K_3 \dfrac{1}{(\beta_0\beta_1)^{3NN_D}} \end{array} \right]$$

$$K_1 = (\gamma_0 NR_C)^{2N^2+NN_D} \left(\frac{1}{\Omega_{SR}^2}\right)^{2N^2} \left(\frac{1}{\Omega_{SD}^2}\right)^{NN_D} (1/N^2!)^2 \times (1/NN_D)(N_0/P)^{2N^2+NN_D}$$

$$K_2 = 2(\gamma_0)^{N^2+NN_D}(NR_C)^{N^2+4NN_D}\left(\frac{1}{\Omega_{SR}^2}\right)^{N^2}\left(\frac{1}{\Omega_{SD}^2}\right)^{NN_D}\left(\frac{1}{\Omega_{RD}^2}\right)^{2NN_D} \times (N_0/P)^{N^2+4NN_D}(1/N^2!)(1/(2NN_D-1)!) \quad (13)$$

$$K_3 = \frac{(\gamma_0)^{3NN_D-1}}{(4P^2)^{3NN_D}(3NN_D-1)!}$$

$$s.t. \quad \beta_0 + \beta_1 + \beta_2 \leq 1$$

$$f_K(\xi) = A\sum_{n=0}^{\infty} \frac{\delta_n \xi^{\sum_{i=1}^{N}\alpha_i+n-1} e^{-\xi/\beta_1}}{\left(\sum_{i=1}^{N}\alpha_i+n-1\right)! \beta_1^{\sum_{i=1}^{N}\alpha_i+n}} U(\xi). \quad (14)$$

$$F_K(\xi) = A\sum_{n=0}^{\infty} \delta_n \frac{\left(\sum_{i=0}^{N}\alpha_i+n\right)^{-1} \xi^{\left(\sum_{i=0}^{N}\alpha_i+n\right)}}{\left(\sum_{i=0}^{N}\alpha_i+n-1\right)! \beta_1^{\sum_{i=0}^{N}\alpha_i+n}} \times {}_1F_1\left(\sum_{i=0}^{N}\alpha_i+n;\ 1+\sum_{i=0}^{N}\alpha_i+n;\ \frac{-\xi}{\beta_1}\right) \quad (18)$$